
\input phyzzx
\def\cL{{\cal L}}
\def\dplus{=\hskip-5pt \raise 0.7pt\hbox{${}_\vert$} ^{\phantom 7}}
\def\dplusup{=\hskip-5.1pt \raise 5.4pt\hbox{${}_\vert$} ^{\phantom 7}}

\REF\us{G. Papadopoulos and P.K. Townsend, Class. Quantum Grav., Vol. 2,
{\bf 3} (1994) 515.}
\REF\wchris{C.M. Hull, G. Papadopoulos and P.K. Townsend, Phys. Lett.
{\bf 316B} (1993) 291.}
\REF\AGF{ L. Alvarez-Gaum{\' e} and D.Z. Freedman, Commun. Math. Phys.
{\bf 91} (1983), 87.}
\REF\geometry{S.J. Gates, C.M. Hull and M. Ro{\v c}ek, Nucl. Phys. {\bf B248}
(1984) 157; C.M. Hull, {\it Super Field Theories} ed H. Lee and G. Kunstatter
(New York: Plenum) (1986)}
\REF\gp {P.S. Howe and G. Papadopoulos, Nucl .Phys. {\bf B289} (1987) 264;
Class. Quantum Grav. {\bf 5} (1988) , 1647.}
\REF\prl{J. A. de Azc{\' a}rraga, J.P. Gauntlett, J.M. Izquierdo and P.K.
Townsend, Phys.
Rev. Lett. {\bf 63} (1989) 2443.}
\REF\Nic {B. de Wit, A.K. Tollst{\'e}n and H. Nicolai, Nucl. Phys. {\bf B392}
(1993) 3.}


\Pubnum{ \vbox{ \hbox{R/94/9} \hbox{DESY 94-092} \hbox{9406015}} }
\pubtype{}
\date{June, 1994}

\titlepage

\title{Massive (p,q)-supersymmetric sigma models revisited}

\author{G. Papadopoulos}
\address{II. Institute for Theoretical Physics
 \break University of Hamburg\break
         Luruper Chaussee 149 \break 22761 Hamburg}
\andauthor{P. K. Townsend}
\address{DAMTP \break Silver Street \break University of Cambridge}

\abstract {We recently obtained the conditions on the couplings of
the general two-dimensional massive sigma-model required by
(p,q)-supersymmetry. Here we compute the Poisson bracket algebra of the
supersymmetry and central Noether charges, and show that the action is
invariant under the automorphism group of this algebra.
Surprisingly, for the (4,4) case the automorphism group is always
a subgroup of $SO(3)$, rather than $SO(4)$.
We also re-analyse the conditions for (2,2) and
(4,4) supersymmetry of the zero torsion models without
assumptions about the central charge matrix.}

\endpage

\pagenumber=2



\def\fff{\vrule width0.5pt height5pt depth1pt}
\def\pp{{{ =\hskip-3.75pt{\fff}}\hskip3.75pt }}

\def\cM {{\cal{M}}}



\chapter{Introduction}

In a recent work [\us] we investigated the restrictions imposed by (p,q)
supersymmetry on the general two-dimensional supersymmetric sigma model
with scalar potential.
Omitting fermions, the action is
$$
 I=\int dxdt\, \Big[ (g+b)_{ij} {\partial_\pp}\phi^i
\partial_= \phi^j  - V(\phi)\big] \ ,
\eqn\aaone
$$
where $\phi$ is a map from the two-dimensional Minkowski space-time,
with light-cone co-ordinates $(x^\pp,x^=)$, into the target manifold $\cM$
with metric $g$. The two-form $b$ is a locally-defined potential
for a globally-defined `torsion' three-form $H$ with components
$H_{ijk}={3\over2}\partial_{[i}b_{jk]}$, and $V$ is the scalar potential.
Since $V$ contains no derivatives its presence requires a coupling
constant $m$ with dimensions of mass; for this reason we refer to models
with $V\ne0$ as `massive' sigma-models.

Here we shall be interested in sigma-models with at least (1,1)
supersymmetry for which the scalar potential takes the form [\wchris]
$$
V= {1\over4}m^2 g^{ij}(u-X)_i(u-X)_j\ ,
\eqn\pot
$$
where $X$ is a (possibly zero) Killing vector field on $\cM$ and $u$ is a
one-form on $\cM$ whose exterior derivative $du$ is determined by $X$ and
$H$ via the formula
$$
 X^k H_{kij}=\partial_{[i} u_{j]}\ .
\eqn\defu
$$
Additional supersymmetries further restrict the scalar potential by
imposing conditions on $X$ and $u$. To find the weakest possible
conditions one must allow for the action of a possible total of $pq$
central charges on the sigma-model fields. These `off-shell' central
charges are associated with $pq$ mutually-commuting Killing vector
fields $\{Z_{I'I}; I=0,1,\dots, p-1,\ I' =0,1,\dots q-1\}$ of which
$X$ is one. Each of them is paired with a one-form $u^{I'I}$
satisfying
$$
Z_{I'I}^k H_{kij}=\partial_{[i} u^{I'I}_{j]}\ ,
\eqn\defugen
$$
in precise analogy with \defu. Requiring that the (1,1)-supersymmetric
sigma-model action be invariant under additional supersymmetry
transformations leads to conditions on the vector-field/one-form pairs
$\{Z_{I'I}, u^{I'I}\}$ which we obtained in our previous work [\us] and
which we summarise in the following section.

We follow this with a presentation of the Noether charges associated with the
supersymmetry and central charge transformations. We then show that closure of
the
algebra of Poisson brackets of these charges requires slightly stronger
conditions conditions than those obtained from consideration of the
transformations
alone, at least if the flat two-dimensional spacetime is assumed to be
Minkowski
spacetime. Specifically, whereas closure of the algebra of transformations was
found to require certain functions to be constants,
closure of the Poisson bracket algebra requires these constants to vanish.

One purpose of this paper is to extend these results to include the
transformations associated with the automorphism group of the
supersymmetry algebra with central charges. By this we mean the subgroup of the
automorphism group of the supersymmetry algebra without central charges that
{\it
leaves invariant the matrix of central charges}. Thus, in our usage, which we
believe to be
standard in the context of supersymmetry, central charges are central not only
in the
supersymmetry algebra but also in the extension of this algebra by its
automorphism
group. For the (2,2) models this group is a subgroup of $SO(2)$ and
for the (4,4) models it is a subgroup of $SO(3)$ (rather than $SO(4)$). We show
that
the automorphism group can be realized in terms of transformations of the sigma
model fields.

Finally, we also presented in
[\us] an analysis of the conditions required by (p,p) supersymmetry in massive
models without torsion, which were first considered by Alvarez-Gaum{\' e} and
Freedman [\AGF]. We argued in section (7) of [\us] that no generality would be
lost if the Killing vector matrix $Z_{I'I}$ of the (p,p) models were assumed to
be diagonal and we deduced the consequences for $V$ under this assumption. Our
argument was based on the fact that $V$ is invariant under $SO(p)\times
SO(q)$ transformations of the matrix $Z_{I'I}$ which, if $p=q$, can be
used to diagonalize it. We implicitly assumed that such an $SO(p)\times
SO(q)$ transformation could be effected by some redefinition of the fermion
fields. Unfortunately, this turns out not to be true so our previous results
for the potentials of the (p,p) models without torsion must be considered
as special cases of a possibly more general result. Another purpose of this
paper is to determine the form of the potential $V$ for these models
without making any assumptions about the matrix $Z_{I'I}$ of Killing vectors.
We present this more general analysis in section (4). Our new result for the
(2,2)
models without torsion is indeed more general than the result of [\us] and
is complete agreement with eq. 50 of [\AGF]\foot{The corresponding result of
[\us] was in apparent agreement with eq. 53 of [\AGF]; there
appears to be a transcription error between eqs. 50 and 53 of [\AGF].}.
We consider this to be a useful check on our results for the general
(p,q) case with torsion.  Our previous conclusions concerning the massive (4,4)
models without torsion remain unchanged but we show here that the automorphism
group of the
supersymmetry algebra of these models is always $SO(3)$ (whereas not all
massive (2,2) models are $SO(2)$ invariant).


\chapter{Massive supersymmetric sigma-models}

In the presence of an off-shell central charge standard (1,1)
superspace methods are inapplicable. The most economical formulation of
the general (p,q)-supersymmetric sigma model is in terms of (1,0)
superfields. The (1,0)-superspace action of the general
(1,1)-supersymmetric model is a functional of the bosonic
(1,0)-superfields $\phi^i$ and the fermionic (1,0)-superfields $\psi^i$,
and takes the form [\wchris]
$$
S=\int\! d^2 x d\theta^+ \big\{
D_+\phi^i\partial_=\phi^j (g_{ij} +b_{ij}) +
i\psi_-^i\nabla^{(-)}_+\psi_-^j \, g_{ij} +
im\, (u_i - X_i)\psi_-^i\big\}\ ,
\eqn\ssaction
$$
where $m$ is a mass parameter and $\nabla^{(\pm)}$ is the covariant
derivative with connection
$$
\Gamma^{(\pm)}_{ij}{}^k = \big\{\matrix{ k\cr ij\cr}\big\}
\pm  H_{ij}{}^k\ ,
\eqn\torsion
$$
i.e. $H_{ijk}$ is the torsion of the connection of $\nabla^{(+)}$. We refer
to [\us] for details of the superspace conventions. The action \ssaction\
is invariant under the superfield transformations
$$
\eqalign{
\delta_\epsilon\phi^i &=-{i\over2}D_+\epsilon_= D_+\phi^i
+\epsilon_=\partial_\dplus\phi^i\cr
\delta_\epsilon\psi_-^i &= -{i\over2}D_+\epsilon_= D_+\psi^i_-
+\epsilon_=\partial_{\dplus}\psi_-^i \ ,\cr}
\eqn\manifest
$$
for $x$-independent (1,0)-superfield $\epsilon_=$. The constant
$(D_+\epsilon_=)|$ is the anticommuting parameter of the manifest
(1,0) supersymmetry\foot{The vertical bar indicates the $\theta=0$
component of a superfield.}. The action \ssaction\ is also invariant under
the transformations
$$
\eqalign{
\delta_\zeta\phi^i &=D_+\zeta \psi_-^i +m\zeta X^i\cr
\delta_\zeta\psi_-^i &=-iD_+\zeta\partial_=\phi^i +
m\zeta\partial_jX^i\psi_-^j}
\eqn\zetatran
$$
for $x$-independent bosonic (1,0)-superfield parameter $\zeta$. The
constant $\zeta |$ is the transformation generated by the
Killing vector $X$ while the anticommuting constant $(D_+\zeta)|$ is the
parameter of (0,1) supersymmetry.

All $(p,q)$-supersymmetric sigma models with $p,q\geq 1$ are special
cases of the (1,1)-supersymmetric model. The additional supersymmetries
simply impose further restrictions on the sigma model couplings
and the geometry of $\cM$. In the massless case these restrictions
are long-established [\geometry, \gp]. For example, an additional $p\, $-1
left-handed supersymmetries requires the existence of $p\, $-1 complex
structures $I_r$ on $\cM$ that are covariantly constant with respect to
the connection $\Gamma^{(+)}$, and that the metric $g$ of $\cM$ be
hermitian with respect to them. In the case that $p=4$, closure of the
algebra of supersymmetry transformations requires in addition that the
complex structures $I_r$ ($r=1,2,3$) obey the algebra of imaginary unit
quaternions. Similarly, an additional $q\, $-1 right-handed supersymmetries
requires the existence of $q\, $-1 complex structures $J_s$ on $\cM$, but in
this case the complex structures $J_s$ are covariantly constant with
respect to the connection $\Gamma^{(-)}$. The metric $g$ must be Hermitian
with respect to all complex structures.

The (1,0)-superfield transformations for extended\foot{Our use of the
adjective `extended' indicates that we exclude the (1,0) and (0,1)
transformations already considered.}
(p,0) and (0,q) transformations are most conveniently  presented in terms
of the `covariantized' fermion variation
$$
\Delta\psi_-^i\equiv \delta\psi_-^i +
\delta\phi^j\Gamma^{(-)}_{jk}{}^i\psi_-^k \ .
\eqn\covdelta
$$
These transformations involve the complex structures $I_r$ and $J_s$ and
several of the Killing-vector/one -form pairs $\{Z_{I'I},
u^{I'I}\}$. For consistency with [\us] we adopt the notation
$$
\eqalign{
Z_{00} =X      &\ \ \  u^{00}=u     \cr
Z_{0r} =Z_r    &\ \ \  u^{0r}=v_r   \cr
Z_{s0} =Y_s    &\ \ \  u^{s0}=w_s   \cr
Z_{sr} =Z_{sr} &\ \ \  u^{sr}=v_{sr}\ .\cr}
\eqn\defaa
$$
The extended (p,0) transformations are
$$
\eqalign{
\delta_\eta\phi^i &=i\eta^r_- I_{r}{}^i{}_j(\phi)D_+\phi^j\cr
\Delta_\eta\psi_-^i &={1\over2}\eta^r_-{\hat I}_r{}^i{}_j(\phi){\cal S}^j
+{im\over2}\eta^r_- (Z_r-v_r)^i \cr}
\eqn\extp
$$
where ${\cal S}^i=0$ is the $\psi^i_-$ field equation and the
(p-1) parameters $\eta_-^r$ are anticommuting constants.
The extended (0,q) transformations are included in\foot{These
transformations can be shown to be equivalent to those of [\us] by
using the various conditions derived in that reference, in particular the
vanishing of the Nijenhuis tensor.}
$$
\eqalign{ \delta_{\kappa}\phi^i &= D_+\kappa^s J_s{}^i{}_j
\psi_-^j +m\kappa^s Y_s^i(\phi)\cr
 \Delta_{\kappa}\psi_-^i &=
iD_+\kappa^s J_s{}^i{}_j\partial_=\phi^j +D_+\kappa^s
J_s^i{}_m H^m{}_{nl}J_s^n{}_j J_s^l{}_k \psi_-^j\psi_-^k +
m\kappa^s \nabla^{(+)}_jY_s{}^i\psi_-^j\cr}
\eqn\extq
$$
where the parameters $\kappa^r$ are $x$-independent bosonic (1,0)
superfields. The (q-1) anticommuting constants
$(D_+\kappa^r)|$ are the parameters of extended
(0,q) supersymmetry. The constants $(\kappa^r)|$ are the
parameters for transformations generated by the Killing vector fields
$Y_r$.

Invariance of the action and closure of the algebra of supersymmetry
transformations requires that each of the Killing vector fields $Z_{I'I}$
leave invariant the complex structures $I_r$ and $J_s$ and the torsion
three-form $H$. It is also required that
$$
Z_{I'I}\cdot u^{J'J}+Z_{J'J}\cdot u^{I'I} =0 \qquad
\cases{I'=J'=0, \quad I=J=0,1,\dots, p-1\cr
I=J, \quad I'=J'=0,1,\dots, q-1} \eqn\afive
$$
Actually, closure of the algebra of supersymmetry transformations was shown
in [\us] to imply only the weaker condition $Z_{I'I}\cdot
u^{J'J}+Z_{J'J}\cdot u^{I'I}={\rm const.}$ but consideration
of the Poisson bracket algebra
of supersymmetry charges, which we discuss in the
following section, shows that
constants some of these constants, those of \afive, must vanish if the
two-dimensional spacetime is Minkowski, as assumed here. For simplicity,
we will take all constants
$$
Z_{I'I}\cdot u^{J'J}+ Z_{J'J}\cdot u^{I'I}=0.
\eqn\afivea
$$

The most important of the remaining restrictions imposed by (p,q)
supersymmetry can now be summarized by the following set of
relations\foot{Correcting a typographical error in [\us].} between the
Killing vector fields $Z_{I'I}$ and their associated one-forms $u^{I'I}$:
$$
\eqalign{ (Z_r + v_r)_i + I_r{}^k{}_i (X+u)_k &= 0
\cr
(Y_s-w_s)_i +  J_s{}^k{}_i(X-u)_k&=0
\cr
(Z_{sr} + v_{sr})_i + I_r{}^k{}_i (Y_s + w_s)_k &=0
\cr
(Z_{sr} - v_{sr})_i + J_s{}^k{}_i (Z_r - v_r)_k &=0 .}
\eqn\aatwo
$$
It was shown in [\us] that the scalar potential $V$ of the general
(p,q)-supersymmetric sigma model can be expressed as the
length of any one of the vectors $Z_{I'I}\pm u^{I'I}$. It follows from
\afivea\ and \aatwo\ that these vectors all have the same
length.


\chapter{ The algebra of charges}

The supersymetry and central charges associated to the symmetries summarized in
the
previous section are most conveniently expressed in terms of the physical
component
fields. For models with at least (1,1) supesymmetry these are
$$
\phi^i=\phi^i| \qquad \lambda_+^i =D_+\phi^i | \qquad
\psi_-^i = \psi_-^i | \ ,
\eqn\compo
$$
where the vertical bar indicates the $\theta^+=0$ component of a superfield.
Performing the $\theta^+$ integration and eliminating auxiliary fields produces
the
component action
$$
\eqalign{
S =\int\! d^2 x\big\{ &\partial_\dplus\phi^i\partial_=\phi^j (g_{ij}+b_{ij})
+ ig_{ij}\lambda_+^i\nabla_=^{(+)}\lambda_+^j -
ig_{ij}\psi_-^i\nabla_\pp^{(-)}\psi_-^j\cr
& -{1\over2}\psi_-^k\psi_-^l\lambda_+^i\lambda_+^j R^{(-)}_{ijkl}
 +m\nabla^{(-)}_i (u -X)_j \lambda_+^i\psi_-^j  -V(\phi) \big\}\ ,}
\eqn\compac
$$
where $V={m^2\over 4} g^{ij} (u-X)_i (u-X)_j$.
The total energy of a sigma model field configuration is
$$
E={1\over 2} \int dx [g_{ij} \partial_t\phi^i
\partial_t\phi^j + g_{ij} \partial_x\phi^i
\partial_x\phi^j + V(\phi) + \ {\rm fermions}\ ]
\eqn\abseven
$$
and the total momentum is
$$
P=\int dx [g_{ij} \partial_t\phi^i \partial_x\phi^j+ {\rm fermions}\ ]\ .
\eqn\abeight
$$
The fermion contributions will not be needed for what
follows so we omit them. Note that the torsion term in the action does not
contribute to the energy. The conserved currents associated with the
Killing vectors $Z^{I'I}$ are
$$
J^{I'I}_\mu = \partial_\mu \phi^i Z_i^{I'I} -
\varepsilon_{\mu}{}^\nu\partial_\nu \phi^i u_i^{I'I}
+ {\rm fermions}\ ,
\eqn\current
$$
where the second term in the current is due to the torsion term in the action.
The corresponding charges are
$$
Q_{I'I}=\int dx [Z^{I'I}{}_i \partial_t\phi^i
+ u^{I'I}{}_i \partial_x\phi^i + {\rm fermions}].
\eqn\abfive
$$
Observe that since $u^{I'I}$
is defined up to the derivative of a scalar the Noether charges are
defined only up the addition of a topological charge. In particular if
the Killing vector field $Z^{I'I}$ vanishes the corresponding charge does not
necessarily vanish but rather becomes a topological charge. These charges must
of
course be taken into account in a determination of the automorphism group of
the
supersymmetry algebra; they will therefore be an important ingredient in our
discussion of this matter in section 5.

The supersymmetry charges of the (p,q) supersymmetric sigma
model can be found by standard manipulations from the
supersymmetry transformations. The results are as
follows. The (1,0)-supersymmetry charge is
$$
S_+=\int dx [g_{ij}
\partial_\pp\phi^i \lambda^j_+-{i\over 2} m (u-X)_i \psi^i_-]\ .
\eqn\abone
$$
The (0,1)-supersymmetry charge is
$$
S_-=\int dx [ i g_{ij} \partial_=\phi^i \psi^j_-
 + {1\over 3} \psi^i_- \psi^j_- \psi^k_-
H_{ijk} + {m\over 2} (X+u)_i \lambda_+^i]\ .
\eqn\abtwo
$$
The extended (p,0)-supersymmetry charges are
$$
S^r_+=\int dx [ I_{r\; ij} (\partial_\pp \phi^i \lambda_+^j-
 i H^i{}_{kl} \lambda^j_+
\lambda^k_+ \lambda^l_+) - i{m\over 2} (v_r-Z_r)_i \psi^i_-]\ .
\eqn\abthree
$$
The extended (0,q)-supersymmetry charges are
$$
S^s_-=\int dx [i J_{s\; ij} \partial_=\phi^i \psi^j_-
+ {1\over 3} H_{mnl} J_s{}^m{}_i J_s{}^n{}_j J_s{}^l{}_k \psi^i_-  \psi^j_-
\psi^k_- +{m\over 2} (Y_s+w_s)_i \lambda_+^i].
\eqn\abfour
$$

To calculate the Poisson bracket algebra of the above charges, one must
first re-express them in terms of the fields $\phi$, $\lambda$,
$\psi$ (and their spatial derivatives) and the corresponding conjugate
momenta which follow in the usual way from the action \compac.  We
omit the details of this step and simply present the result of the subsequent
calculation of the Poisson Brackets.
Firstly,
$$
\{S_+,S_+\}= 2 (E+P), \quad \{S^r_+,S^s_+\}= 2 \delta_{rs} (E+P),
 \quad \{S_+,S^r_+\}=0 \ .
\eqn\absix
$$
One does not expect central charges to appear in these anticommutators
because their presence is forbidden by two-dimensional Lorentz invariance.
However,
the calculation shows that there are Lorentz non-invariant central
charges proportional to the volume of space; a typical such charge is
$\int\! dx\, X\cdot u\, $. As we remarked earlier, closure of the algebra of
supersymmetry transformations implies that the integrand is constant, so the
charge
is infinite if space is infinite, as it is for two-dimensional Minkowski
spacetime.
Under these circumstances these constants must vanish\foot{They need not vanish
if
space is compact. Analogous central charges are important in the context
of supersymmetric extended objects in toroidal spacetimes [\prl].}
and the Poisson bracket
algebra of the (p,0)-supersymmetry charges is as given in \absix. Under the
same
circumstances the algebra of Poisson brackets of the (0,q)-supersymmetry
charges is
$$
\{S_-,S_-\}=2(E-P),
\quad \{S^r_-,S^s_-\}= 2 \delta_{rs} (E-P), \quad \{S_-,S^s_-\}=0\ .
\eqn\abnine
$$

Lorentz-invariant central charges can appear in the Poisson brackets of the
(p,0)-supersym\-metry charges with the (0,q) ones, and indeed we find that
$$
\eqalign{
\{S_+,S_-\}&=m Q_{oo}, \quad \{S_+,S_-^s\}= m Q_{so},
\cr
\{S_-,S_+^r\}&= m Q_{or}, \quad \{S_+^r,S_-^s\}= m Q_{sr}\ .}
\eqn\abten
$$
Finally, the Poisson brackets of $Q_{I'I}$ with any supersymmetry charge
vanishes.
Thus we have now obtained the Noether charges of the general massive
(p,q)-supersymmetric sigma-model and verified that they realize the algebra of
(p,q) supersymmetry with central charges.


\chapter{ Automorphism symmetries}

All of the transformations summarized in section 2 have
parameters that are $x$-indepen\-dent (1,0) superfields except those of the
extended (p,0) transformations for which the parameters $\eta_-^r$ were
(anticommuting) constants. If $m=0$ this restriction is unnecessary and
the action remains invariant if $\eta_-^r$ is also an $x$-independent
superfield. The symmetry with parameters $(D_+\eta_-^r)|$ is an $SO(N)$
rotation of the fermions $(D_+\phi^i)|$, where $N=2$ if $p=2$ and
$N=3$ if $p=4$. There is also an $SO(M)$
invariance of the form
$$
\delta_\xi\psi_-^i = \xi^s R_s{}^i{}_j \psi_-^j\ ,
\eqn\caone
$$
for constant parameters $\xi_s$, provided that the tensors $R_s$ satisfy
$$
(R_s)_{(ij)}=0 \qquad \nabla_i^{(-)} (R_s)_{kl}=0\qquad
\eqn\cbone
$$
and
$$
[R_r,R_s] =  \sum_t\varepsilon_{rst} R_t\ .
\eqn\cfour
$$
The commutator of \caone\ with the (0,1)-supersymmetry transformations
yields an extended (0,q)-supersymmetry transformation provided that
$$
R_s = J_s\ ,
\eqn\cfoura
$$
which implies both \cbone\ and \cfour\ and so replaces them. The other
commutators yield no further conditions. From \cfoura\ it is seen that $M=2$ if
$q=2$
and $M=3$ if $q=4$. The $SO(N)\times SO(M)$ symmetry of the massless
sigma-model may
clearly be identified with a subgroup of the $SO(p)\times SO(q)$ automorphism
group
of the supersymmetry algebra without central charges.

We now turn to the $m\ne0$ case. For reasons that will become apparent
below we combine the rotation \caone\ with the extended
(p,0) transformations to form the transformations
$$
\eqalign{
\delta_{(\eta,\xi)}\phi^i &=i\eta_-^r I_r{}^i{}_j D_+\phi^j\cr
\Delta_{(\eta,\xi)}\psi^i_- &= \eta_-^r \hat I_r{}^i{}_j{\cal S}^j
+ {im\over2}\eta_-^r(Z_r-v_r)^i + i\xi^s J_s{}^i{}_j\psi_-^j \ ,}
\eqn\cfourb
$$
which we shall call the {\it new} extended (p,0) transformations. The
commutator on $\phi$ of the new extended (p,0) transformations among
themselves does not produce any conditions not already found from the
$m=0$ case. Omitting terms proportional to field equations, the
commutator on $\psi_-$ is
$$
\eqalign{
[\delta_{(\eta,\xi)},\delta_{(\eta',\xi')}]\psi_-^i &=
-\Gamma^{(-)}_{jk}{}^i\big\{[\delta_{(\eta,\xi)},
\delta_{(\eta',\xi')}]\phi^j\big\}\psi_-^k -
2i\eta'_-{}^r\eta_-^r\nabla_{\dplus}^{(-)}\psi_-^i \cr
& -2\xi'{}^s\xi^t\varepsilon_{stw}J_w{}^i{}_j\psi_-^j
+{m\over2}\big(\xi^s\eta'_-{}^r-\xi'{}^s\eta_-{}^r\big)
\big(J_s(Z_r-v_r)\big)^i .}
\eqn\cfourc
$$
The right hand side can be identified with known transformations provided
that
$$
\xi^r = D_+\eta_-^r \qquad r=1,\dots,{\rm min}(p-1,q-1)\ ,
\eqn\cfourd
$$
and
$$
(v_r -Z_r)_i +(J_r)_{ij}(u-X)^j=0 \ .
\eqn\cthree
$$
Indeed, using \cfourd\ the new extended (p,0) transformations \cfourb\ can be
rewritten in terms of the single $x$-independent superfield $\eta_-^r$
and, using \cthree, the on-shell commutator of these transformations is found
to be
$$
\eqalign{
[\delta_\eta,\delta_{\eta'}]&\psi_-^i =
-\Gamma^{(-)}_{jk}{}^i\big\{[\delta_\eta,\delta_{\eta'}]\phi^j\big\}\psi_-^k\cr
&- 2i\big(\eta'_-{}^r\eta_-^r\nabla_{\dplus}^{(-)}\psi_-^i
-{i\over 2}D_+(\eta'_-{}^r\eta_-^r)\nabla_+^{(-)}\psi_-^i\big)\cr
& -{1\over2}\varepsilon_{rst}\big[m(D_+\eta'_-{}^r\eta_-^s +\eta'_-{}^r
D_+\eta_-^s) (Z-v)^i_t + 4 D_+\eta'_-{}^r D_+\eta_-^s J_t{}^i{}_j \psi_-^j
\big]\ .}
\eqn\cfoure
$$
The explicit connection term on the right hand side cancels the connection
terms
implicit in the covariant derivatives. One can then see that the commutator
closes on (1,0) and new extended (p,0) transformations (up to field equations).

The commutator of the (0,1) transformations with the new extended (p,0) ones
is
$$
\eqalign{
[\delta_\eta,\delta_\zeta]\phi^i = {1\over2}D_+\zeta\eta_-^r &(\hat
I_r- I_r)^i{}_j {\cal S}^j -im\eta_-^rD_+\phi^k({\cal L}_XI_r)^i{}_k\cr
&-im\eta_-^rD_+\zeta Z_r^i + iD_+\zeta D_+\eta_-^r J_r{}^i{}j\psi_-^j\ .}
\eqn\cfive
$$
For simplicity we shall again consider on-shell closure, which means
that we may ignore the first term on the right hand side (we refer the
reader to [\us] for a more complete discussion of this point). The second term
vanishes because $X$ is holomorphic with respect to all complex structures. For
{\it constant} parameters $\eta_-^r$ the last term vanishes while the
third term can interpreted as a central charge transformation. This is
not possible when the parameters $\eta_-^r$ are $x$-independent
superfields rather than constants, and in this case the commutator produces
a potentially-new symmetry for which the variation of $\phi$ is
$$
\delta \phi^i =  iD_+(\eta_-^rD_+\zeta) J_r{}^i{}_j\psi_-^j -im
(\eta_-^rD_+\zeta)Z_r^i\ .
\eqn\csix
$$
We can identify this transformation as that of an extended (0,q)
supersymmetry transformations \extq\ provided that
$$
\qquad Z_r = -Y_r \qquad r=1,\dots, {\rm min}(p-1,q-1)\ .
\eqn\csevena
$$
It turns out that the commutator on $\psi_-$ closes in the same way without the
need of any further conditions. Clearly, at most a diagonal $SO\big({\rm
min}(N,M)\big)$ subgroup of the $SO(N)\times SO(M)$ symmetry of the massless
model can be realized in the massive case.

Using the condition ${\cal L}_{Y_s}I_r=0$ required for closure of the
supersymmetry algebra, the on-shell commutator on $\phi$ of the extended
(0,q) transformations \extq\ with the new extended (p,0) ones is
$$
\eqalign{
[\delta_\eta,\delta_\kappa]\phi^i = &i D_+\kappa_-^r D_+\eta_-^s
\varepsilon_{rst}J_t^i{}_j\psi_-^j -iD_+\kappa_-^r D_+\eta_-^r\psi_-^i\cr
 &+ imD_+\kappa_-^r\eta_-^s Z_{sr}^i\ .}
\eqn\csevenaa
$$
For $p=q=2$ the right hand side can be identified with known transformations
provided that
$$
Z_{11}\equiv T = X\ .
\eqn\csevenab
$$
Similarly, for $p=q=4$ the right hand side can be identified with known
transformations provided that
$$
Z_{sr}= \delta_{sr} X - \sum_t\varepsilon_{srt}Y_t\ .
\eqn\ceight
$$
We shall not trouble the reader with the complications of the $p\ne q$ cases
except
to say that if we take $p<q$ then the results are essentially the same as those
of
the (p,p) model. No additional conditions arise from consideration of the
commutator
on $\psi_-$. Finally, the commutators of the new extended (p,0) transformations
with
the (1,0) supersymmetry transformations close without the need of any further
conditions. We have still to consider whether the action is invariant under the
new
extended (p,0) transformations; a calculation shows that the action is
invariant as
a consequence of the conditions required for closure of the algebra.


\chapter{Automorphism algebra for p=q}

We have just established the conditions for invariance of the sigma model under
additional bosonic symmetries which we have called `automorphism' symmetries.
It
is clear from the way they were found that they are indeed automorphisms of the
supersymmetry transformations in the sense that a commutator of an
`automorphism'
transformation with a supersymmetry transformation yields a further
supersymmetry
transformation. The connection with the automorphism group of the Poisson
bracket
algebra of supersymmetry charges is less clear, however, particularly in view
of
the fact that the Killing vector matrix $Z_{I'I}$ is not necessarily
proportional
to the central charge matrix because the latter may contain topological
charges. It
is also not so clear why only an $SO(3)$ automorphism can be realized in the
massive
(4,4) models. These points will be adressed in the course of this and the
following
section. Here we shall show for the general $p=q$ sigma model that the the
group
realized by the automorphism transformations indeed coincides with the
automorphism
group of the Poisson bracket algebra of the supersymmetry charges. We also
explain
why $SO(4)$ cannot be realized in (4,4) models.

To discuss the (2,2) models it is convenient to define
$$
(Y_1, w^1)= (Y,w),\quad  (Z_1,v^1)=(Z,v),\quad
(Z_{11},v^{11})=(T,n)\ .
\eqn\ceightaa
$$
We showed in the previous section that these models are invariant
under an $SO(2)$ symmetry provided that
$$
(v -Z)_i +(J)_{ij}(u-X)^j=0
\eqn\ceighta
$$
and
$$
Z = - Y \qquad T=X\ .
\eqn\ceightb
$$
These conditions are of course additional to those of \aatwo. The combined set
of
equations implies that
$$
v=-w \qquad n=u
\eqn\ceightc
$$
and
$$
\eqalign{
(w-Y)_i + J^j{}_i(u-X)_j &=0 \cr
(w+Y)_i - I^j{}_i(u+X)_j &=0\ .}
\eqn\ceightd
$$
The {\it independent} conditions are \ceightb, \ceightc, and \ceightd.
The significance of equations \ceightb\ and \ceightc\ is that they ensure that
the central charge matrix is invariant under an $SO(2)$
 subgroup of the $SO(2)\times SO(2)$ automorphism group of
 the supersymmetry algebra without central charges.
Equation \ceightb\ is obviously necessary for this, but the necessity of
\ceightc\ is perhaps less obvious. To
see why it is necessary consider the Noether charges $Q_{I'I}$
given in \abfive: when $Z=-Y$ and $T=X$ we find that
$$
\eqalign{ Q_{01} &= -Q_{10} +\  {\rm surface\ term}\cr
Q_{00} &=Q_{11} +\  {\rm surface\ term}\ , }
\eqn\ceightf
$$
where the surface terms can be interpreted as topological
charges. These topological charges must also vanish if the central charge
matrix
is to be $SO(2)$ invariant, and the conditions of \ceightc\ ensure that this
occurs, i.e. that the central charge matrix $Q$ takes the form
$$
Q = \pmatrix{ Q_X & -Q_Y\cr
Q_Y & Q_X}\ ,
\eqn\ceightg
$$
where $Q_X\equiv Q_{00}$ and $Q_Y\equiv Q_{01}$.

We now turn to the (4,4) models. We have found that these models are invariant
under an $SO(3)$ symmetry provided that the conditions
$$
(v_r -Z_r)_i +(J_r)_{ij}(u-X)^j=0\ ,
\eqn\canine
$$
and
$$
\eqalign{
Z_r &= - Y_r\cr
Z_{sr} &= \delta_{sr} X - \sum_t\varepsilon_{srt}Y_t \ ,}
\eqn\cnine
$$
hold. Again, these are in addition to those of \aatwo.
The combined set of equations implies
$$
v_r=-w_r \qquad v_{sr}=\delta_{sr} u -\sum_t\varepsilon_{srt}w_t
\eqn\cninea
$$
and
$$
\eqalign{
(w_r -Y_r)_i + J_r{}^j{}_i (u-X)_j &=0\cr
(w_r +Y_r)_i - I_r{}^j{}_i (u+X)_j &=0\ .}
\eqn\cnineb
$$
The {\it independent}  equations are those of \cnine, \cninea, and \cnineb.
The equations \cnine\ and \cninea\ ensure that the central charge matrix $Q$
takes the form
$$
Q = \pmatrix{ Q_X & -Q_1 & -Q_2 & -Q_3 \cr
Q_1 & Q_X & -Q_3 & Q_2  \cr
Q_2 & Q_3 & Q_X & -Q_1 \cr
Q_3 & -Q_2 & Q_1 & Q_X }\ ,
\eqn\cninec
$$
where $Q_X\equiv Q_{00}$ and $Q_r\equiv Q_{r0}$.
This matrix is a sum of a multiple of the identity matrix and a
self-dual matrix. This is the general form for a matrix that is
invariant under an antiself-dual $SO(3)$ subgroup of the diagonal $SO(4)$.

The above analysis can also be carried out directly at the level of the
Poisson brackets. The Noether charges associated with the automorphism
symmetries are
$$
A^r={1\over 2} \int dx \big(-i I_r{}_{ij}\lambda_+^i \lambda_+^j+i
J_r{}_{ij}\psi_-^i\psi_-^j  \big).
\eqn\abeleven
$$
Using the conditions derived in section 4 one finds that the
Poisson brackets of these charges with themselves
and the other charges of the supersymmetry
algebra are, for the $p=q=2$ model,
$$
\eqalign{ \{&A,A\}=0, \qquad \{A,S_+\}=S^1_+, \qquad \{A,S^1_+\}=-S_+,
\cr
\{&A,S_-\}=S_-^1, \qquad \{A,S_-^1\}=-S_-, \qquad \{A,H\}=0,
\cr
\{&A,P\}=0, \qquad \{A,Q\}=0\ ,}
\eqn\abtwelvef
$$
where $A\equiv A^1$ and $Q\equiv Q^{11}$, and, for
the $p=q=4$ model,
$$
\eqalign{
\{&A^r,A^s\}=-2 \varepsilon_{rst} A^t, \qquad \{A^r, S_+\}=
S_+^r
\cr
 \{&A^r, S_+^s\}= - \delta_{rs} S_+ - \varepsilon_{rst} S_+^t
\cr
\{&A^r, S_-\}= S_-^r, \qquad \{A^r, S_-^s\}=- \delta_{rs} S_-
- \varepsilon_{rst} S_-^t
\cr
\{&A^r, H\}=0, \qquad \{A^r, P\}=0, \qquad \{A^r, Q^{I'I}\}=0 \ .}
\eqn\abtwelve
$$
As expected, in both cases the automorphism charges $A^r$ transform
the supersymmetry charges amongst
themselves but leave the Hamiltonian, $H$, the momentum, $P$, and the central
charges, $Q_{I'I}$, invariant.

Given the fact that an $SO(2)$ symmetry can be realized for certain (2,2)
models,
one might have expected to be able to realize an $SO(4)$ symmetry for some
(4,4)
models. It seems, however, that at most an $SO(3)$ subgroup of $SO(4)$ can be
realized\foot{A related phenomenon occurs in the work of [\Nic] on the coupling
of
$N=4$ three-dimensional supersymmetric sigma-models to $N=4$ supergravity; the
$SO(4)$ invariance of the pure supergravity action is maintained in the
matter-coupled action by virtue of the fact that the supergravity fields couple
to
{\it two} sigma-models,  each of which contributes an $SO(3)$ factor.}. The
relevant $SO(4)$ group for the massive models is the {\it diagonal} subgroup of
the
$SO(4)_L\times SO(4)_R$ automorphism group acting on the left (L) and right (R)
handed supercharges of the massless sigma model. It can be shown that only one
of
the two $SO(3)$ subgroups  of this $SO(4)$ can  be realised by transformations
on
the fields of a massive (4,4) supersymmetric sigma model.

We shall first show that for the massless (4,4) supersymmetric
sigma models, one can realise only an $SO(3)_L\times SO(3)_R$
subgroup of the $SO(4)_L\times SO(4)_R$ automorphism group of the supersymmetry
algebra.  For this, it is sufficient to examine
the (4,0) sector of the algebra since the proof for the other
sector is identical. We first observe that the
diagonal $SO(4)_L$ acts on the supersymmetry charges $S^I_+$
via its fundamental representation $D$;
$$
D(T_{KL})^J{}_I=\delta_{KI} \delta_L{}^J-\delta_{LI} \delta_K{}^J \eqn\cten
$$
where $K,L,I,J=0,\dots, 3$ and $T_{KL}\equiv-T_{LK}$ is a basis in the
Lie algebra of $SO(4)_L$. The self-dual and antiself-dual parts of this
representation are
$$
D^{(\pm)}_{KL}=D(T_{KL})\pm {1\over 2} \varepsilon_{KL}{}^{MN} D(T_{MN})\ .
\eqn\celeven
$$
They form four-dimensional representations of two commuting $SO(3)$ subgroups
of
$SO(4)_L$ as can be seen by definining
$$
D^{(\pm)}_r=D^{(\pm)}_{0r}
\eqn\ctwelve
$$
and observing that
$$
D^{(\pm)}_r D^{(\pm)}_s=-\delta_{rs}\pm \sum_t\varepsilon_{rst} D^{(\pm)}_t
\eqn\cthirteen
$$
and
$$
[D^{(+)}_r,D^{(-)}_s]=0\ .
\eqn\cthirteena
$$
We shall denote by $SO(3)_L^+$ and $SO(3)_L^-$ the corresponding $SO(3)$
subgroups
of $SO(4)_L$.  The matrices $D^{(\pm)}$ form a four
dimensional representation of $SO(3)_L^{\pm}$
and neither $SO(3)$ factor of the $SO(4)_L$ automorphism
group leaves invariant any of the supercharges $S_+^I$. To see this, let
us assume that there is one, say $S_{+}\equiv S_+^0$, which is invariant
under the transformation $D^{(\pm)}=
\xi^r D^{(\pm)}_r$ for infinitesimal parameter $\xi$, i.e.  $D^{(\pm)}S_+=0$.
We then observe that
$(D^{(\pm)})^2=-(\xi)^2 {\bf 1}$ and thus $(\xi)^2S_+=0$ which
implies that $\xi=0$.

Recall now from section 4, that a massless (4,4) supersymmetric sigma model
is invariant under {\it independent} rotations of the
fermions $\lambda_+$ and $\psi_-$.  The
corresponding Noether charges are
$$
A^r_L=-{i\over 2} \int dx I_{rij} \lambda^i_+ \lambda^j_+, \qquad
A^s_R={i\over 2} \int dx J_{sij} \psi^i_- \psi^j_-\ .
\eqn\cthirteenb
$$
After some computation, it is straightforward to show that
$$
\{A^r_L, S_+^I\}={D_r}^{(-)}{}^I{}_J S_+^J\ .
\eqn\cthirteenc
$$
{}From this it is clear that one can realise the $SO(3)_L^-$ subgroup
of $SO(4)_L$ with rotations on the fermions $\lambda_+$
induced by the complex structures
$I_r$ of the sigma model manifold.

To realise the $SO(3)_L^+$ subgroup of $SO(4)_L$, one may consider
introducing some further rotations on the fermion fields $\lambda_+$. For
this, one needs a set of $(1,1)$ tensors, $F_r$ say, that differ from $I_r$,
but
invariance of the action and closure of the algebra will require that
$F_r=I_r$,
so no realisation of $SO(3)_L^+$ is possible by a rotation of the fermion
fields.
Alternatively, one might try to realise the  $SO_L^+(3)$ by rotating the bosons
instead of the fermions. This would involve consideration of transformations of
the form
$$  \eqalign{
\delta\phi^i&=\xi^r k_r
\cr
\delta\lambda_+^i&=\xi^r \partial_jk^i_r \lambda^j_+
\ ,}
\eqn\cfourteen
$$
where $\xi^r$ are parameters and $k_r$ are vector fields which must be Killing
and leave invariant $H$ in order to be invariances. In order to qualify as
automorphism symmetries they would also have to rotate the
complex structures $I_r$ into themselves
(${\cal L}_{k_r}I_s=\sum_t \varepsilon_{rst} I_t$).
But because the vector fields $k_r$
are Killing they leave invariant the particular supersymmetry charge
$S_+\equiv S^0_+$ (given in section 3) and hence cannot realise the
$SO(3)_L^+$ subgroup of the automorphism group $SO(4)_L$.

The above arguments can be repeated for $SO(4)_R$, to prove that
only its subgroup $SO(3)_R^-$ can be realised by transformations on the fields.
Therefore only the subgroup $SO(3)_L^-\times SO(3)_R^-$ of the
automorphism group $SO(4)_L\times SO(4)_R$ of the supersymmetry
algebra without central charges can be realised by symmetries in the
massless sigma model.

In the massive ($m\not=0$) case, closure of the
algebra of transformations (given in section 4) requires that the parameters of
the left and right rotations of the fermions $\lambda_+$ and $\psi_-$ be
the same and the resulting Noether charge is $A^r=A^r_L+A_R^r$.
Combining this fact with the above discussion for the massless model we
conclude that for the massive model only the diagonal subgroup $SO(3)^-$ of
$SO(3)_L^-\times SO(3)_R^-$ can be realised.

To conclude, we remark that the $SO(3)$ group that
leaves invariant the central charge matrix $Q$ of \cninec,
and hence the automorphism group of the (4,4) supersymmetry
algebra with central charges,
is the $SO(3)^-$.  To prove this, we note that
$$
Q=Q_X\, {\bf 1}-\sum_rQ_r D_r^{(+)}
\eqn\cfourteena
$$
and then that \cthirteena\ implies $[D^{(-)}_r, Q]=0$.

\chapter{Non-chiral models revisited}

We shall now present some more detailed results for the special
case of the non-chiral models with extended supersymmetry, i.e. the (2,2)
and (4,4) models without torsion. For vanishing torsion it is possible to
set $J_r=I_r$ and here we shall consider only this case.

We first consider the (2,2) models; recall that $(Y_1, w^1)= (Y,w)$,
$(Z_1,v^1)=(Z,v)$, and $(Z_{11},v^{11})=(T,n)$. Since $H=0$ we have
$u^{I'I}= da^{I'I}$, i.e.
$$
u=da \qquad v=db \qquad w=dc \qquad n= de
\eqn\AGFcase
$$
for locally-defined functions $a$, $b$, $c$ and $e$. From \aatwo\ we
now deduce that
$$
\eqalign{ (Y-Z)_i &= I^k{}_i \partial_k(a+e) \cr
(X+T)_i &= I^k{}_i\partial_k (b-c)\cr
(X-T)_i &=I^k{}_i (Z+Y)_k\cr
\partial_i(a-e) &= I^k{}_i\partial_k (c+b)\ . }
\eqn\AGFtwo
$$
In addition we know that $X$, $Y$, $Z$ and $T$ are Killing
vector fields that are holomorphic ($\cL_XI=\cL_YI=\cL_ZI=\cL_TI=0$) with
respect to a covariantly constant ($\nabla I=0$) complex structure.
Given certain conditions on the global structure\foot{A sufficient
condition is that $\cM$ be compact and simply connected.}
of the target manifold $\cM$, any such
holomorphic Killing vector field, $k$, can be expressed in terms of an
associated real Killing potential $U$ as $k_i=I^j{}_i \partial_j
U$. Thus, from the first two equations of \AGFtwo\ we may identify
$(a+e)$ and $(b-c)$ as the Killing potentials of $(Y-Z)$ and $(X+T)$,
respectively. Similarly, the other two independent linear combinations
may also be written as
$$
(Y+Z)_i = I^k{}_i \partial_k \alpha \qquad  (X-T)_i =
I^k{}_i \partial_k \beta\ ,
\eqn\kilpot
$$
where the scalars $\alpha$ and $\beta$ are the Killing potentials. It
follows directly from \AGFtwo\ and \kilpot\ that
$$
(Y+Z)_i = -\partial_i\beta \qquad (X-T)_i =\partial_i\alpha \ .
\eqn\kilpotb
$$
A solution to \kilpot\ and \kilpotb\ is\foot{A sign error in a similar
analysis in section (7) of [\us] led to the incorrect equation $T=-X$.}
$$
Z=-Y \qquad T=X
\eqn\AGFc
$$
with $\alpha$ and $\beta$ constant. If $\cM$ is either irreducible or
compact and simply connected then this solution is unique. Either of
these conditions is sufficient to prove uniqueness although neither is
necessary. It can also be shown that the form of the scalar potential is
unchanged if the general solution is used when the solution \AGFc\ fails to be
unique. For simplicity, we shall assume here that $\cM$ is such that \AGFc\ is
the only solution of \kilpot\ and \kilpotb.
{}From the first two equations of \AGFtwo\ we then find that
$$
X_i = I^k{}_i \partial_k
\left({c-b\over2}\right)\qquad Y_i = I^k{}_i
\partial_k\left({a+e\over2}\right)\ ,
\eqn\AGFd
$$
from which we may identify the Killing potentials of the two independent
Killing vector fields. Let $\gamma$ be the Killing potential of $Y$; then
$$
a+e= 2\gamma + \ {\rm constant}\ .
\eqn\AGFe
$$
The last of eqs. \AGFtwo\ implies that ${1\over2}(a-e)$ is the real part of
a holomorphic function, i.e.
$$
a-e = 2(h+ \bar h)\ ,
\eqn\AGFf
$$
where $h$ is holomorphic. Eliminating $e$ from \AGFe\ and \AGFf\ we find
that
$$
a = \gamma + (h+\bar h) + \ {\rm constant}
\eqn\AGFg
$$
and hence that
$$
\eqalign{
|da|^2 &= g^{ij}
\big[\partial_i\gamma + \partial_i(h+\bar h)\big]
\big[\partial_j\gamma +  \partial_j(h+\bar h)\big]\cr
&= |Y|^2 + |d(h+\bar h)|^2 +
2g^{ij}\partial_i\gamma\partial_j(h+\bar h)\ .}
\eqn\AGFh
$$
Now,
$$
\eqalign{
g^{ij}\partial_i\gamma\partial_j (h+\bar h) &=
g^{ij}\partial_i\gamma\partial_j (a-e)\cr
&= I^{ki}\partial_i\gamma \partial_k (b+c)\cr
&= Y^k\partial_k (b+c)\cr
&= {\rm constant}\ ,}
\eqn\AGFi
$$
since $d(b+c)$ is $Y$-invariant. Thus
$$
\eqalign{
V &= {1\over4}m^2\big(|X|^2 + |da|^2\big) \cr
&= {1\over4}m^2\big(|X|^2 + |Y|^2 + |d(h+\bar h)|^2\big)\ ,}
\eqn\AGFj
$$
in agreement with eq. 50 of [\AGF].

Observe that the restriction \AGFc\ on the Killing vector fields is such
that the the vector-valued matrix $Z_{I'I}$ takes the $SO(2)$-invariant form
$$
Z_{I'I}= \mu\delta_{I'I} + \nu\varepsilon_{I'I}\ .
\eqn\ccharges
$$
As explained in section 4, this is necessary but
not sufficient for the $SO(2)$ invariance of the action. For this one
also needs $c=-b$ and $e=a$, in which case
$$
X_i =-I^k{}_i\partial_k c\ , \qquad Y_i = I^k{}_i\partial_k a \ ,
$$
and $h=0$. Thus, $SO(2)$ invariance requires the superpotential, $h$,
to vanish.

We turn now to the (4,4) models. Since $H=0$ we have
$$
u=da \qquad v_r = db_r \qquad w_r= dc_r \qquad v_{sr} = de_{sr}\ .
\eqn\xfoura
$$
Defining
$$
T_r\equiv Z_{rr} \qquad \qquad (r=1,2,3)\ ,
\eqn\xfourb
$$
we deduce from \aatwo\ that
$$
\eqalign{ (Y_r-Z_r)_i &= I_r{}^k{}_i \partial_k(a+e_{rr}) \qquad (r=1,2,3)\cr
(X+T_r)_i &= I_r{}^k{}_i\partial_k (b_r-c_r)\qquad (r=1,2,3)\cr
(X-T_r)_i &=I_r{}^k{}_i (Z_r+Y_r)_k \qquad (r=1,2,3)\cr
\partial_i(a-e_{rr}) &= I_r{}^k{}_i\partial_k (c_r+b_r)\qquad (r=1,2,3)\ , }
\eqn\xfourc
$$
which is the (4,4) analogue of \AGFtwo . Now, however, \aatwo\ implies
the further conditions
$$
\eqalign{
(Z_{[rs]})_i&=-{1\over2} (I_sI_r)^k{}_i \partial_k (e_{rr}+e_{ss}),
\qquad (r\ne s),
\cr
\partial_ie_{[rs]}&=-{1\over2} (I_sI_r)^k{}_i (T_r+T_s)_k,
\qquad (r\ne s),
\cr
(Z_{(rs)})_i&=-{1\over2} (I_sI_r)^k{}_i  (T_s-T_r),\qquad (r\ne s),
\cr
\partial_ie_{(rs)}&=-{1\over2} (I_sI_r)^k{}_i \partial_k (e_{rr}-e_{ss})
\qquad (r\ne s).}
\eqn\xfourd
$$
The third condition in \xfourd\ is similar to the third  equation in \AGFtwo,
and applying the same arguments as in that case we can deduce that
$$
Z_{(rs)}=0, \qquad T_s-T_r=0, \qquad (r\ne s)\ .
\eqn\xfourda
$$
A consequence of this is that
$$
e_{(rs)}=0, \qquad e_{rr}-e_{ss}=0, \qquad (r\ne s)\ ,
\eqn\xfoure
$$
(up to constants), and the fourth equation in \xfourd\ is automatically
satisfied.
Thus the three functions $\{e_{rr}; r=1,2,3\}$ are actually the same
function, and similarly for the three Killing vectors $T_r$, so it is
convenient
to define
$$
e_{rr}= e, \qquad T_r= T \qquad (r=1,2,3)\ .
\eqn\xfourf
$$
We can also write
$$
\eqalign{
Z_{[rs]} &= \sum_t \varepsilon_{rst}\, W_t\cr
de_{[rs]} &= \sum_t \varepsilon_{rst}\, df_t\ .}
\eqn\xfourf
$$
The residual information contained in eqs. \xfourd\ can now be expressed
in terms of the functions $e$ and $f_r$ and the Killing vector fields $T$
and $W_r$ as
$$
\eqalign{
T_i &= I_r{}^k{}_i\partial_k f_r \qquad (r=1,2,3)\cr
(W_r)_i &= -I_r{}^k{}_i\partial_k e \qquad (r=1,2,3)\ ,}
\eqn\xfourg
$$
while using \xfourf\ to simplify \xfourc\ we deduce that
$$
\eqalign{ (Y_r-Z_r)_i &= I_r{}^k{}_i \partial_k(a+e) \qquad (r=1,2,3)\cr
(X+T)_i &= I_r{}^k{}_i\partial_k (b_r-c_r)\qquad (r=1,2,3)\cr
(X-T) &=I_r{}^k{}_i (Z_r+Y_r)_k \qquad (r=1,2,3)\cr
\partial_i(a-e) &= I_r{}^k{}_i\partial_k (c_r+b_r) \qquad (r=1,2,3)\ .}
\eqn\xfourh
$$
For a given value of $r$ the equations \xfourh\ are precisely
those of the (2,2) case, \AGFtwo. With the same assumptions as before
about the global structure of $\cM$ we deduce that
$$
T=X\qquad \qquad Z_r= -Y_r\qquad (r=1,2,3)\ .
\eqn\xfouri
$$
The remaining three equations of \xfourh\ combined with those of \xfourg\
then reduce to
$$
\eqalign{
X_i &={1\over2}I_r{}^k{}_i \partial_k (b_r-c_r) \qquad (r=1,2,3)\cr
(Y_r)_i &= {1\over2}I_r{}^k{}_i \partial_k (a+e) \cr
(W_r)_i &= -I_r{}^k{}_i \partial_k e \cr
\partial_i(a-e) &= I_r{}^k{}_i\partial_k (c_r+b_r) \qquad (r=1,2,3)}
\eqn\xfourj
$$
which imply that
$$
(Y_r+W_r)_i = -{1\over2}\partial_i(c_r+b_r)\ .
\eqn\xfourk
$$
Hence the argument that previously led to the conclusion that
$Z_r=-Y_r$ now leads to
$$
W_r=-Y_r \qquad c_r= -b_r \ ,
\eqn\xfourl
$$
and using this to simplify \xfourj\ yields
$$
\eqalign{
X_i &= I_r{}^k{}_i\partial_k b_r \qquad (r=1,2,3)\cr
(Y_r)_i &= I_r{}^k{}_i\partial_k a\ . }
\eqn\xfourm
$$
The first of these equations allows us to identify the functions $b_r$ as
the three Killing potentials of the triholomorphic Killing vector $X$.
The second shows that the Killing vectors $Y_r$ are holomorphic with
respect to $I_r$ with Killing potential $a$ but we also know [\us]
that they must be triholomorphic, i.e. there exist functions $m_{rs}$
such that
$$
(Y_r)_i = I_s{}^k{}_i\partial_k m_{rs}\qquad (r,s=1,2,3)\ .
\eqn\xfourn
$$
Combining this with the equation for $Y_r$ in \xfourm\ yields
$$
\partial_i a = (I_sI_r)^k{}_i \partial_k m_{rs}\qquad (r,s =1,2,3)
\eqn\xfouro
$$
and this implies that
$$
\partial_i a= I_r{}^k{}_i \partial_k m_r\ ,
\eqn\xfourp
$$
where $m_t$ is defined by $dm_{[rs]} = \sum_t \varepsilon_{rst}\, dm_t$.
This is the condition that $a$ be a triholomorphic function. Substituting
this result into the second of eqs. \xfourm\ we see that $Y_r = -dm_r$
and then, by the previous argument for target spaces of the assumed
global structure, $Y_r=0$. Thus we have now shown that
$$
da =0
\eqn\xfourq
$$
and hence that the potential $V$ is simply the length of the
triholomorphic Killing vector field $X$.

{}From the results of section 3, we now see that the (4,4) models without
torsion are $SO(3)$ invariant. The matrix of Killing vector fields is actually
$SO(4)$ invariant but, as explained in section 3, this does not necessarily
imply that the central charge matrix is $SO(4)$ invariant because of the
possibility of topological charges. These topological charges vanish
identically (i.e. for {\it all} sigma model field configurations) if and only
the functions $b_r$ are all constant. But in this case $X=0$ and so the
scalar potential $V$ vanishes. Thus, the central charge matrix for massive
(4,4)
models without torsion is actually only $SO(3)$ invariant and this corresponds
precisely to the group realized by the `automorphism' transformations of the
fields.


\noindent{\bf Acknowledgements:} Discussions with C.M. Hull and H. Nicolai are
gratefully acknowledged. G.P. was funded by a grant from the European
Union.

\refout


\end